# Inverse design of two-dimensional materials with invertible neural networks

Victor Fung,[1,*] Jiaxin Zhang,[2,*] Guoxiang Hu,[3] P. Ganesh,[1] Bobby G. Sumpter[1]

[1] Center for Nanophase Materials Sciences, Oak Ridge National Laboratory, Oak Ridge, Tennessee 37831, United States
[2] Computer Science and Mathematics Division, Oak Ridge National Laboratory, Oak Ridge, Tennessee 37831, United States
[3] Department of Chemistry and Biochemistry, Queens College of the City University of New York, Queens, New York 11367, United States
*E-mail: fungv@ornl.gov, zhangj@ornl.gov

**Abstract**

The ability to readily design novel materials with chosen functional properties on-demand represents a next frontier in materials discovery. However, thoroughly and efficiently sampling the entire design space in a computationally tractable manner remains a highly challenging task. To tackle this problem, we propose an inverse design framework (MatDesINNe) utilizing invertible neural networks which can map both forward and reverse processes between the design space and target property. This approach can be used to generate materials candidates for a designated property, thereby satisfying the highly sought-after goal of inverse design. We then apply this framework to the task of band gap engineering in two-dimensional materials, starting with $MoS_2$. Within the design space encompassing six degrees of freedom in applied tensile, compressive and shear strain plus an external electric field, we show the framework can generate novel, high fidelity, and diverse candidates with near-chemical accuracy. We extend this generative capability further to provide insights regarding metal-insulator transition, important for memristive neuromorphic applications among others, in $MoS_2$ which is not otherwise possible with brute force screening. This approach is general and can be directly extended to other materials and their corresponding design spaces and target properties.

# Introduction

In materials discovery problems, it is desirable to select and test candidates which hold the most promise for satisfying a particular functional target, while maintaining as broad a diversity in the search space as possible. To this end, a data-driven approach is often used to meet these needs, whereby materials are first rapidly screened via high-throughput experimentation or computational modeling to identify potential candidates.[1,2,3] However, the vastness of the accessible chemical search spaces can present serious challenges for current experimental or even computational methods due to the significant evaluation time and resource demands. Machine learning provides promising solutions to this problem by providing a cheaper surrogate for the computational calculations, or by producing new candidates which have a specified target property.[4,5,6,7,8,9] The latter approach may involve the use of generative models, which allows for sampling within a continuous chemical or materials latent space which can map to unique and undiscovered materials.[10,11,12,13,14] Although more challenging to implement than discriminative models, generative modeling is highly appealing for its potential to realize the "inverse design" of materials and to efficiently "close the loop" between modelling and experiments.[4,15,16,17,18]

In general inverse problems, given a forward process $y = f(x)$, the goal is to then find a suitable inverse model $x = f^{-1}(y)$ to map the reverse process. In the context of materials discovery, the forward mapping from materials design parameters to target property can take the form of experimental measurements or computational calculations, such as density functional theory (DFT). However, the reverse process from target property to design parameters cannot be obtained directly via the same methods but can be inferred with machine learning. One such approach uses variational autoencoders[19] (VAEs), where the encoder and decoder models learn to approximate inverse solutions upon convergence. Instead, we propose to use invertible neural networks[20] (INNs) and conditional INNs[21] (cINNs) where a single model can be trained on a forward process and the exact inverse solution can then be obtained for free. The intrinsic invertibility of INNs offers potential advantages in stability and performance[20,21,22,23] over VAEs and the popular generative adversarial networks[24] (GANs), which suffers from difficulties in training due to mode collapse.

In this work, we leverage the INN architecture to solve materials design problems and develop a framework utilizing these models to generate high-quality materials candidates with the targeted properties, outlined in Figure 1. Starting with a given materials design space, we begin by generating training data sampling within this space using DFT. We use the data to train the invertible neural network to obtain forward and reverse mappings. The trained network is then evaluated in the reverse direction to generate samples given a target property. We note the generated samples may not directly yield sufficiently high-fidelity candidates within chemical accuracy of the target. We therefore include additional (1) down-selection based on fitness criteria and (2) optimization to localize generated samples to the exact solutions using the initialized samples from the INN. In this work, the fitness criteria for down-selection limits samples to those which are close to the desired property and with parameters within the training data distribution, but this can be further expanded to include additional criteria such as experimental feasibility as needed. Selected samples are then optimized by gradient descent with automatic differentiation.[25] For down-selection and optimization, the forward mode of the INN can be conveniently used as the property prediction surrogate and provide gradients via backpropagation. These final optimized samples may be further validated by DFT, or if the performance is sufficiently high, may be omitted and the generated samples then analyzed directly. We show our framework, MatDesINNe (Materials Design with Invertible Neural Networks), can successfully provide inverse materials design with high accuracy accuracy.

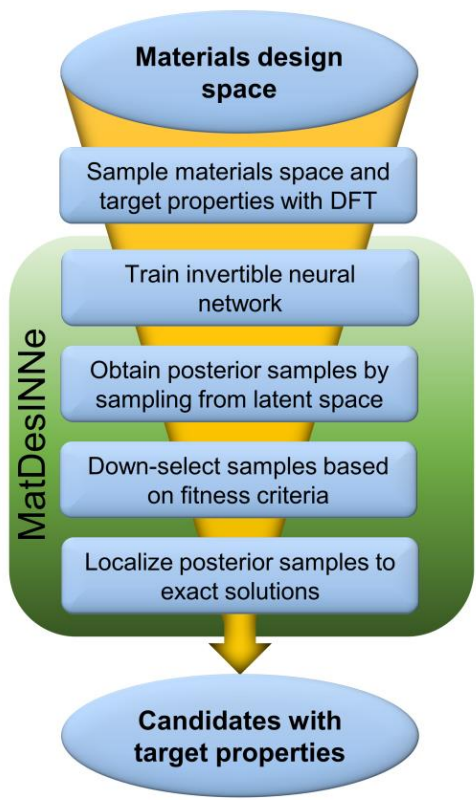

**Figure 1:** Workflow of inverse materials discovery procedure. Starting with a specified design space, training data is obtained with DFT which is then fed to train the MatDesINNe framework using invertible neural networks. Samples are generated, down-selected and localized to ensure high quality candidates. An optional validation step with DFT ensures the candidates have the intended target properties.

We apply MatDesINNe to the band gap engineering problem for monolayer $MoS_2$, the archetypical 2D dichalcogenide, where the design parameters are applied strain and external electric field and the target property is the electronic band gap ($E_g$). Strain can been applied to 2D materials to monotonically tune their band gaps, which can be experimentally accomplished via many methods including bending or stretching the substrate, thermal expansion, or through the application of local stress from an atomic force microscope tip.[26, 27, 28, 29, 30] Similarly, electric field can also be used to further modulate the band gap due to band-bending, and provides an additional degree of freedom which is not constrained by the elastic strength of the material.[31, 32] The ability to tune the band gap freely allows the material to be designed for a target application, including in photocatalysis, electronics, sensors, and neuromorphic devices.[26, 33, 34, 35, 36]

## Results

To generate the training data, we performed approximately 11,000 DFT calculations by sampling over the entire range of the design space (Figure 2a). In this work, we represent applied strain as deviations in the equilibrium lattice constants a, b, c, α, β, γ, with a sampling range of 20% above and below the equilibrium. While this may exceed realistic achievable ranges in some cases, we focus first on achieving an accurate mapping of the forward and reverse process from the full design space, and experimental limitations can be applied later at the down-selection stage. In addition to strain, an external electric field is applied in a range from -1 to 1 V/Å. The DFT-calculated band gap at equilibrium is 1.97 eV, and we find most samples lie below this value under the applied strain and electric field (Figure 2b). Most notably, the vast majority of cases result in an insulator-metal transition (MIT) to a band gap of zero, and this data imbalance can present significant challenges for generative models. We also illustrate the full design space in two dimensions using the Uniform Manifold Approximation and Projection (UMAP) method[37] (Figure 2c), which shows no clear decision boundary between low $E_g$ and high $E_g$ states, especially in the 0-1 eV region where substantial intermixing of states can be observed. Therefore, the generative model used must also have a high degree of fidelity to avoid large errors in the target property of generated candidates.

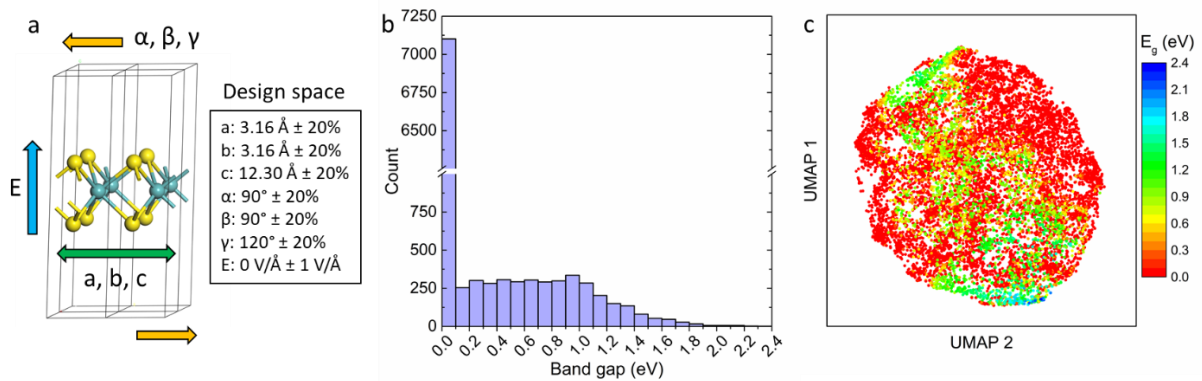

**Figure 2:** (a) $MoS_2$ model and design parameter space specification, which can be categorized as tensile/compressive (green arrow), shear (orange arrow) and electric field (blue arrow). (b) Distribution of DFT-computed band gaps within the sampled design space for monolayer $MoS_2$. (c) UMAP embedding of sampled design space with DFT-computed band gaps

To determine whether our approach succeeds for this type of challenging inverse problem, we first train and subsequently validate our model by comparing the band gap of generated samples with the surrogate model. We also compare our models with several other inverse-mapping methods in order to calibrate performance: (1) mixture density methods

(MDN)[38] and (2) conditional variational autoencoders (cVAE)[39]. For our models we implemented the base (3) invertible neural networks (INN)[20] and (4) conditional invertible neural networks (cINN)[21] as well as the additional steps included in the MatDesINNe framework as (5) MatDesINNe-INN and (6) MatDesINNe- cINN. The results are compiled in Table 1, where 10,000 samples are obtained with each model and validated against the surrogate to reveal their relative performance. We find all models performed adequately well for the $E_g = 0$ case, an unsurprising result, as statistically most samples will have zero band gaps. However, for non-zero cases such as $E_g = 0.5$ and $E_g = 1.0$ the performance of the baseline models drops dramatically, far too low for use in any materials design situation. With invertible neural networks, cINN maintains an appreciable performance of ~0.2 eV for both non-zero gap cases, though the error still remains higher than ideal. When down-selection and optimization are then performed on cINN posterior samples, the error is further reduced to near-zero for all target cases. The significant improvement in performance is made possible due to the effective initialization of posterior samples with cINN, whereas posterior samples provided by INN are unable to be localized to near-zero error due to the comparatively poorer initialization. Similarly, MDN and cVAE methods tested here would also be expected to fail to localize effectively for the same reason.

**Table 1:** Surrogate-validated performance and speed of tested models for 10000 samples given three targets: $E_g = 0$, 0.5 and 1 eV.

| Method | $E_g$=0 eV | | $E_g$=0.5 eV | | $E_g$=1.0 eV | |
|---|---|---|---|---|---|---|
| | MAE (eV) | Time (s) | MAE (eV) | Time (s) | MAE (eV) | Time (s) |
| MDN | 0.184 | 5.255 | 0.421 | 5.288 | 0.840 | 4.934 |
| cVAE | 0.064 | 5.246 | 0.461 | 5.532 | 0.973 | 5.711 |
| INN | 0.038 | 5.912 | 0.527 | 5.598 | 0.835 | 5.891 |
| cINN | 0.063 | 5.833 | 0.219 | 6.026 | 0.193 | 5.591 |
| MatDesINNe-INN | 0.038 | 28.035 | 0.512 | 32.096 | 0.321 | 33.598 |
| MatDesINNe-cINN | **0.020** | 27.882 | **0.013** | 30.903 | **0.015** | 30.953 |
| DFT | N/A | 8.43E+05 | N/A | 9.60E+06 | N/A | 1.03E+07 |

    In terms of computational cost, all machine learning models required less than a minute to generate 10,000 samples on a single standard CPU. MatDesINNe requires slightly more time due to the additional localization step but provides substantially improved performance. The speeds provided for the models can be considered to be essentially on-the-fly for the generative task, which can be further increased with using GPUs. Regardless of method, the computation times are negligible compared to the DFT calculations: to generate 10,000 samples with a band gap of $1 \pm 0.1$ eV with DFT would require $1.03 * 10^7$ seconds or over five orders of magnitudes longer time than MatDesINNe due to both the intrinsic DFT evaluation time and the low statistical probability for finding target band gap. The speedup provided here with inverse

learning can be considered as a lower bound for general problems, as more complex properties and higher levels of theory will necessitate much longer DFT evaluation times.

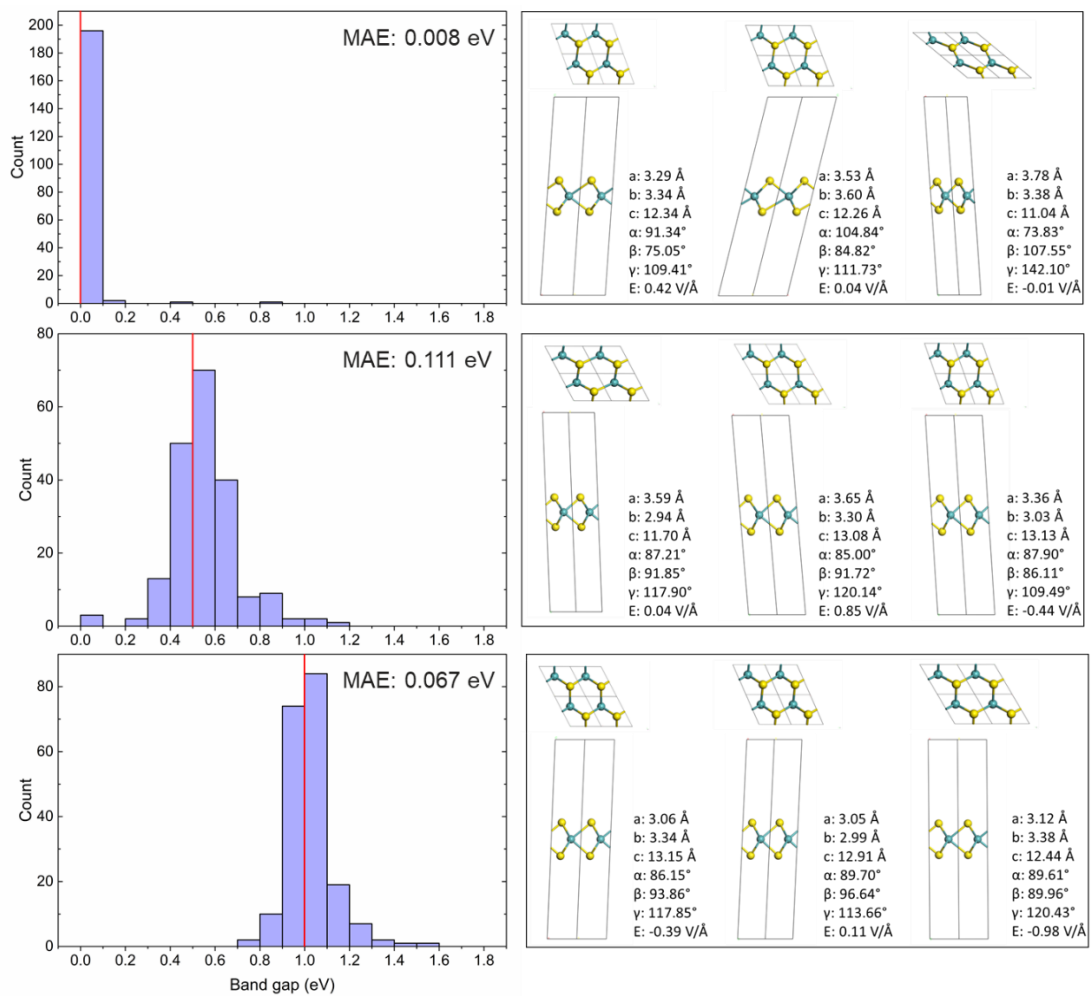

**Figure 3:** DFT-validated performance of generated candidates for three targets: $E_g = 0$, 0.5 and 1 eV (shown by the position of the red line). Example structures are shown for each target.

We then validate the best-performing model, MatDesINNe-cINN with DFT calculations to find the absolute performance relative to DFT. The DFT band gaps for each of the three targets are shown in Figure 3 for 200 samples, showing an excellent performance with a MAE of 0.1 eV or lower. This accuracy is only slightly higher than usual experimental error bars for band gap measurements,[40] which is a usual a point of comparison when discussing chemical accuracy. While our model is trained on DFT band gaps which contain its own error bars, we expect a model expressive enough to fit well to the DFT energies to do similarly well for another more accurate quantum chemistry method. A low incidence of outliers is also observed; only 6 out of 200 samples, or 3%, have non-zero gaps for $E_g=0$, and less than 2% of samples have zero band gaps for $E_g=0.5$ and $E_g=1$. The absolute performance here is limited by the accuracy of the surrogate model used for gradient-based optimization, hence the discrepancy between DFT-validated performance and earlier surrogate-validated performance in Table 1. By improving the accuracy of the surrogate model and thereby allowing the optimized samples to be closer to the true DFT value, we anticipate the DFT-validated performance can be further improved.

In addition to the high fidelity of the generated candidates with this approach, we find they adequately cover the distribution of the original training data and while maintaining a high degree of diversity in the design parameter space. To demonstrate this, distributions of the training data and generated data are compared side-by-side for $E_g = 0, 1$ in Figure 4, and plotted with respect to average in-plane (a, b, γ) and out-of-plane (c, α, β) strain for ease of visualization. The distributions for both $E_g = 0$ and $E_g = 1$ are shown to match quite well with the training data and not localized to a specific region in the design space. This example illustrates another strength of this approach compared to methods such as conditional GANs which often struggle with maintaining a high sample diversity.[21]

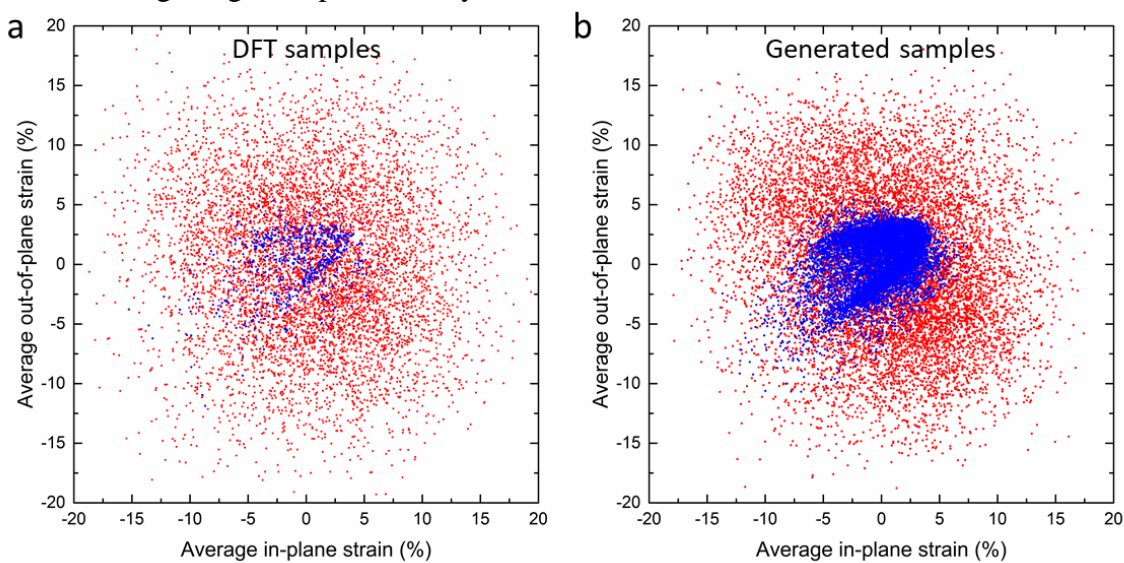

**Figure 4:** Distributions of average in-plane strain vs. out-of-plane strain for (a) DFT training data and (b) generated data cases. For each case, red denotes $E_g = 0$ eV samples, while blue denotes the $E_g = 1 \pm 0.1$ eV band gap samples.

With a sufficiently accurate and expressive generative model as demonstrated here, the problem of generating specific design parameters with a target band gap becomes trivial. We next expand upon these capabilities by investigating the overall parameter space as mapped out by the model. For example, we can group the strain parameters into two categories: tensile/compressive (a, b, c) and shear (α, β, γ) and plot their overall distributions in Figure 5. Here, we find the distribution in average shear strain is fairly normal, while it is significantly more skewed for tensile/compressive strain. It is readily apparent $E_g$ is far more sensitive to shear strain than tensile/compressive strain, where few samples exist for $E_g = 1$ when average shear strain is over 5% in either direction. When viewed with respect to the absolute strain (Figure 5c, d), zero gap and non-zero gap samples can be better distinguished, with a rough correlation where higher absolute strain leads to greater probability for finding $E_g = 0$ materials, especially for the shear strain case. Meanwhile, $E_g = 0.5$ and 1 share much of the same strain space, with the exception of the tensile strain region (positive values) where $E_g = 1$ drops off quickly but $E_g = 0.5$ persist up to approximately 17% in tensile strain. These analyses can help provide design principles and provide guidance into regions of the strain space for further sampling.

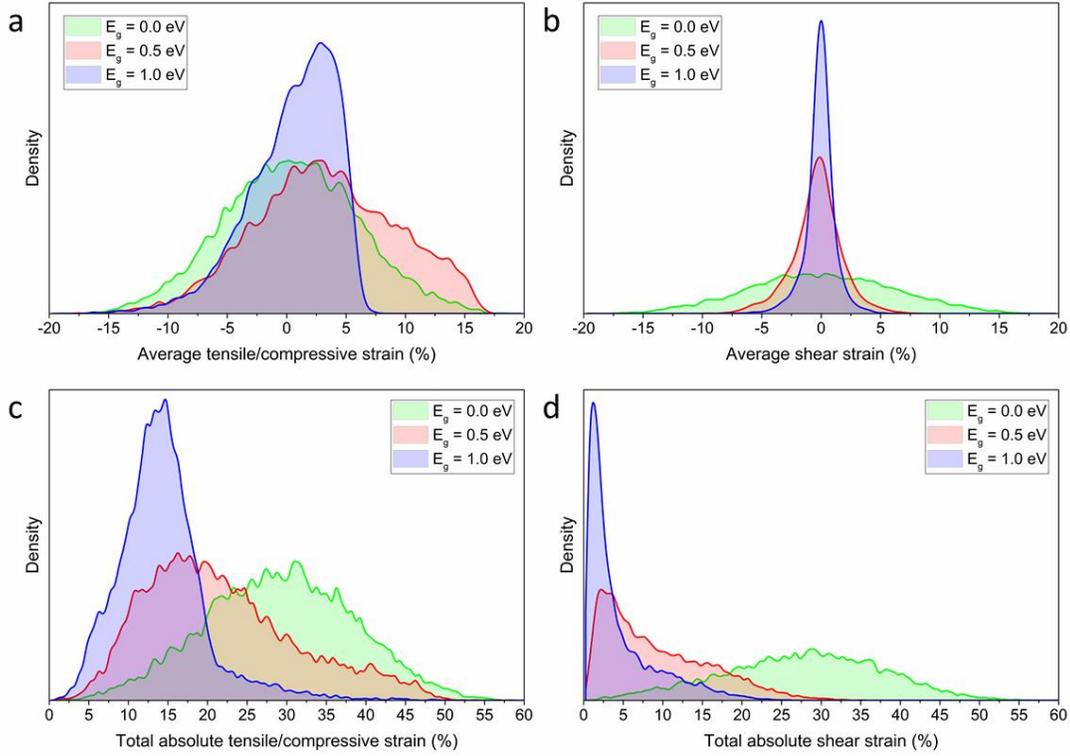

**Figure 5:** Distributions of generated data by (a) average tensile/compressive strain, (b) average shear strain, (c) average absolute tensile/compressive strain and (d) average absolute shear strain for three targets: $E_g = 0$, 0.5 and 1 eV.

For a more targeted application, we propose to use the generated samples to probe the metal insulator transition of $MoS_2$, a key property in neuromorphic devices. One useful insight gained by generative modeling is revealing regions in the design parameter space where MIT occurs with minor perturbations. To illustrate this, we use UMAP to reduce the 7-dimensional space into two dimensions for two situations, the MIT between $E_g = 0$ and 0.5 (Figure 6a) and the MIT between $E_g = 0$ and 1 (Figure 6b). In terms of global structure, we see two relatively distinct regions in the reduced dimensional space for both cases, though the $E_g = 0$ and 1 case shows a much clearer decision boundary. This is an expected behavior consistent with the observation suggested previously in Figure 5 in which a high $E_g$ correlates with low shear strain and vice versa with little overlap. Meanwhile, locally, there are many regions in the reduced dimensional space where zero and nonzero band gaps coexist, even for the case of $E_g = 0$ and 1. As UMAP tends to locally group entities which are similar in the full input dimension, an overlap in points with zero and nonzero gaps suggest a transition between these two states only requires a minor change in the applied strain with two examples shown in the zoomed-in insets for Figure 6. In the example for the $E_g$=0 and 0.5 case, MIT occurs from a 7% tensile strain in the y-axis and a 0.82 V/Å change in electric field. For the $E_g$=0 and 1 case, a 6% tensile strain in the y-axis and 7% compressive strain in the z-axis is needed. This can highlight potentially useful regions in strain space where MIT can be easily induced, allowing for fast and energy-efficient switching. Alternatively, if the goal is to prevent the occurrence of MIT, it is then desirable to select regions in the strain space where no $E_g$=0 cases can be found in the vicinity.

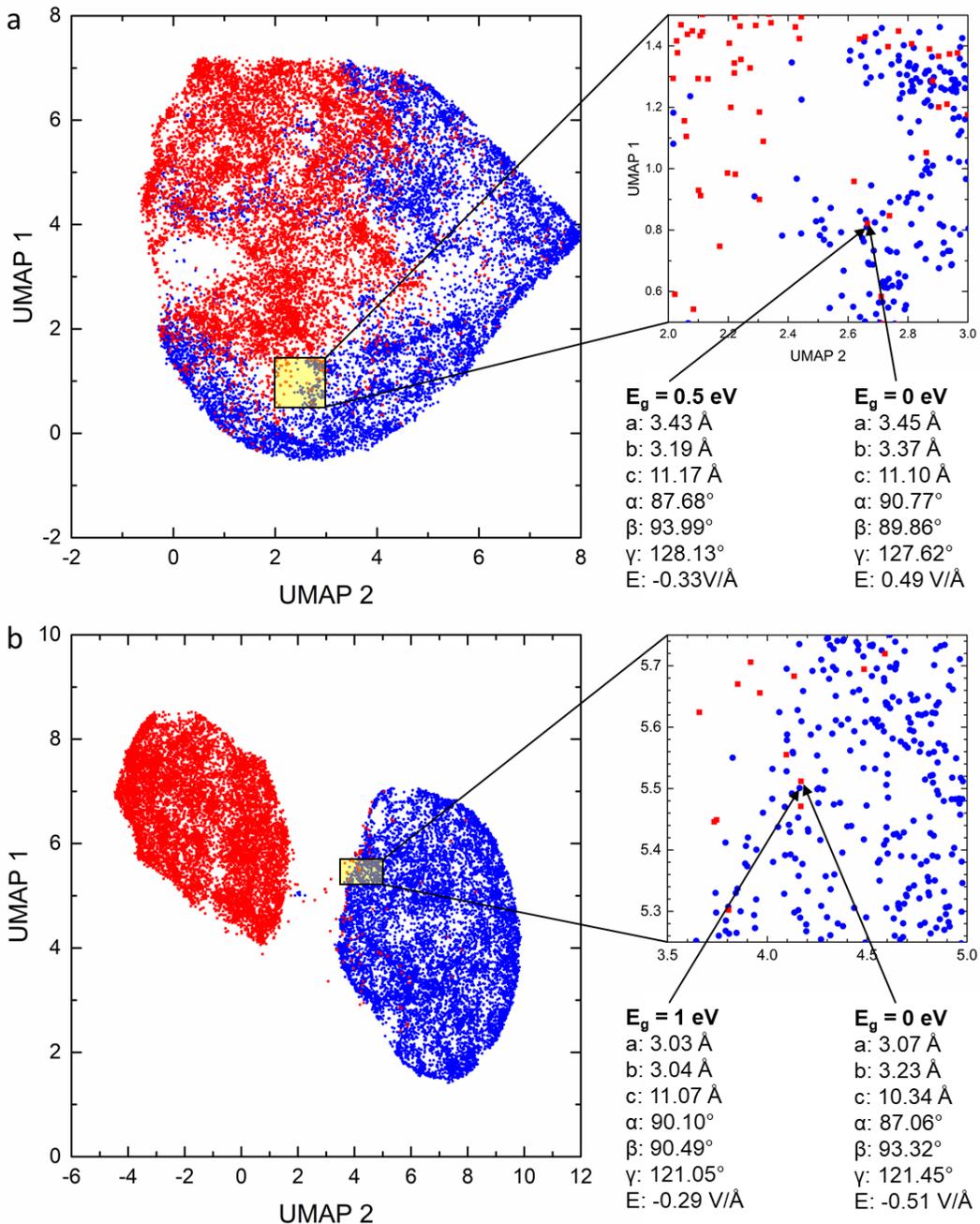

**Figure 6:** UMAP embedding of strain parameters for (a) $E_g = 0.5$ and (b) $E_g = 1$ eV with $E_g = 0$ eV samples. Red denotes $E_g = 0$ eV samples, while blue denotes the nonzero band gap samples. Selected regions (in yellow) are zoomed in to highlight specific examples where MIT occurs.

## Discussion and conclusion

INNs exhibit the attractive property of intrinsic invertibility and serves as a promising approach for solving inverse design problems. In our materials design framework MatDesINNe, we train INNs and run evaluations in the reverse mode to generate samples given a target property. To ensure samples have target properties within chemical accuracy, we further incorporate down-selection and localization via optimization on the generated samples. We apply the framework to engineering band gaps in monolayer $MoS_2$ via applied strain and electric field. We find our approach outperforms other baseline state-of-the-art methods and succeeds at the conditional generative task despite the challenging nature of the problem, with a target MAE of around 0.1 eV or lower. This approach is then used provide high-fidelity and diverse candidates at several orders of magnitude lower computational cost than direct screening with DFT. Using this model, we then generate large quantities of candidate materials to explore the design space and obtain useful design principles or insights for further sampling. Finally, we show how our model can help tackle complex problems such as the MIT in $MoS_2$ by densely populating the design space for two target states and identifying potential regions where fast switching of states can be achieved.

Moreover, our approach is materials and applications agnostic and can be applied for general materials design tasks, provided well-defined design parameters and target properties(s) are chosen and available for training. We have made this code available in an open-source repository. In the current work, we focus on lattice strain and did not include additional atomic degrees of freedom as design parameters, which has been included in recent studies using variational autoencoders (VAEs) and generative adversarial networks (GANs).[41, 42, 43, 44] These additionals dimensions can be included in our current model as well without much difficulty, and INNs have been shown to work for high-dimensional problems.[21] However, generative methods involving atomic structure remain hampered by a lack of an invertible crystal representation which is furthermore rotationally and translationally invariant.[11] We leave the problem of a generative model for atomic structures for the future when such a representation is developed.

We anticipate a materials design framework such as MatDesINNe will provide both theoretical insights as shown in this work as well as offer a means of integrating computation with experiments. The traditional approach of linear discovery via make, measure, model remains severely limited by the large amount of variables, complex coupled/competing underlying phenomena, and slow process times. Methods which can navigate the design space in a rational fashion to select experiments, as well as learn the outcome of these experiments can therefore greatly improve efficiency and introduce autonomous guidance. By generating high fidelity and diverse samples on-the-fly with chemical accuracy, MatDesINNe can satisfy many of these requirements and be incorporated into the autonomous experimentation process.

## Methods

**Density functional theory** Lattice constants and angles a, b, c, α, β, γ were sampled within a range of 20% deviation from the equilibrium crystal parameters: a=b=3.16 Å, c=12.30 Å, α=β=90°, γ=120°. An external electric field is also applied in the z-direction from -1 to 1 V/Å. A total of 10799 structures were generated and band gaps obtained using density functional theory (DFT). The DFT calculations were performed with the Vienna Ab Initio Simulation Package (VASP)[45, 46]. The Perdew-Burke-Ernzerhof (PBE)[47] functional within the generalized-gradient approximation (GGA) was used for electron exchange and correlation energies. The projector-augmented wave method was used to describe the electron-core interaction[45, 48]. A kinetic energy cutoff of 500 eV was used. All calculations were performed with spin polarization. The Brillouin zone was sampled using a Monkhorst-Pack scheme with a 9x9x1 grid.[49]

**Machine learning**

*Inverse problem specification.* Typically, a mathematical or physical model is developed to describe how measured observations $\mathbf{y} \in \mathbb{R}^M$ arise from the hidden parameters $\mathbf{x} \in \mathbb{R}^D$ to yield such a mapping $\mathbf{y} = \Omega(\mathbf{x})$. To completely capture all possible inverse solutions given observed measurements, a proper inverse model should enable the estimation of the full posterior distribution p(x|y) of hidden parameters x conditioned on an observation y.

*Invertible neural networks.* A recent study[21] showed that the invertible neural networks (INNs) can first be trained in the forward pass and then used in the reverse mode to sample from p(x|y) for any specific y. This is achieved by adding a latent variable $\mathbf{z} \in \mathbb{R}^K$ (K=D-m), which encodes the inherent information loss in the forward process. In other words, the latent variable z drawn from a Gaussian distribution $p(\mathbf{z}) = \mathcal{N}(0, I_K)$ is able to encode the intrinsic information about x that is not contained in y. To this end, an augmented inverse problem is formulated based on such a *bijective* mapping, as shown in Figure 7:

$$\mathbf{x} = h(\mathbf{y}_a; \phi) = h(\mathbf{y}, \mathbf{z}; \phi), \quad \mathbf{z} \sim p(\mathbf{z})$$

Where h is a deterministic function of y and z, parametrized on an INN with parameter $\phi$. Forward training optimizers the mapping $\mathbf{x} \rightarrow \mathbf{y}_a = [\mathbf{y}, \mathbf{z}]$ and implicitly determines the inverse mapping $\mathbf{x} = h(\mathbf{y}, \mathbf{z})$. In the context of INNs, the posterior distribution p(x|y) is represented by the deterministic function $\mathbf{x} = h(\mathbf{y}, \mathbf{z})$ that transforms the known probability distribution p(z) to parameter x-space, conditional on measurements y. Thus, given a chosen observation y* with the learned h, we can obtain the posterior samples $x_k$ which follows the posterior distribution p(x|y*) via a transformation $x_k = h(y^*, z_k)$ with prior samples drawn from $z_k \sim p(z)$.

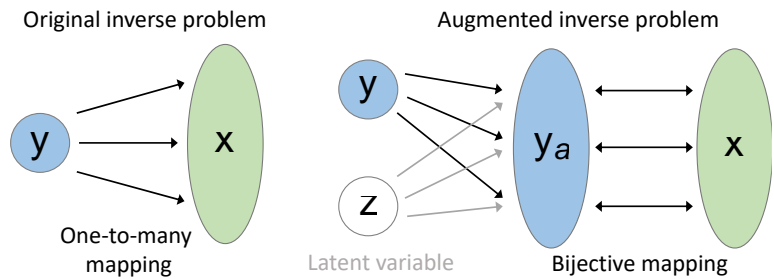

**Figure 7**. The original inverse problem is often ill-posed due to the one-to-many mapping. An augmented inverse problem is formulated based on bijective mapping with introducing additional latent random variable z.

The invertible architecture allows us to simultaneously learn the model h(y,z; $\phi$) of the inverse process jointly with a model f(x; $\phi$) which approximates the true forward process $\Omega(x)$:

$$[\mathbf{y}, \mathbf{z}] = f(\mathbf{x}; \phi) = [f_\mathbf{y}(\mathbf{x}; \phi), f_\mathbf{z}(\mathbf{x}; \phi)] = h^{-1}(\mathbf{y}, \mathbf{z}; \phi)$$

Where $f_\mathbf{y}(\mathbf{x}; \phi) \approx \Omega(\mathbf{x})$, model f and h share the same parameters $\phi$ in a single invertible neural network. Therefore, our approximated posterior model $\hat{p}(\mathbf{x}|\mathbf{y})$ is built into the invertible neural network representation as

$$\hat{p}(\mathbf{x} = h(\mathbf{y}, \mathbf{z}; \phi)|\mathbf{y}) = p(\mathbf{z})/|J_\mathbf{x}|, \quad J_\mathbf{x} = \det\left(\frac{\partial h(\mathbf{y}, \mathbf{z}; \phi)}{\partial [\mathbf{y}, \mathbf{z}]}\right)$$

Where Jx is the Jacobian determinant that can be efficiently computed by using affine coupling blocks.[50]

In this work use an invertible architecture with affine coupling layers.[20] A forward L2 loss is defined where $y_t$ is the true output:

$$L_y = \|y - y_t\|_2^2$$

A backward MMD loss $L_z$ is used to fit the probability distribution of latent variable p(z) to a standard Gaussian distribution. The total loss is then defined with weighting factors $\lambda y$ and $\lambda z$:

$$L = \lambda_y L_y + \lambda_z L_z$$

*Conditional invertible neural networks.* Instead of training an invertible neural network to predict y and x with additional latent variable z, the conditional invertible neural network[21] transforms x directly to a latent representation z conditional on the observation y. This is achieved by using y as an additional input to each affine coupling layer in both forward and backward processes. The model is then trained with a maximum likelihood loss:

$$L = \frac{1}{2} \cdot z^2 - \log |\det J_{x \to z}|$$

*Conditional variational autoencoders.* The conditional variational autoencoder[39] uses the evidence lower bound and encodes the x into a Gaussian distributed random latent variable z conditioned on y. The forward training process utilizes the L2 loss to achieve a good reconstruction of the original input x while the backward process solves x which is decoded from random samples that are drawn from z conditioned on y. The loss function is defined as:

$$L = \alpha \cdot (x - \hat{x})^2 - \frac{1}{2} \cdot \beta \cdot (1 + \log \sigma_z - \mu_z^2 - \sigma_z)$$

*Mixture density networks.* Mixture density networks[38] directly model the inverse problem, with y as the input and predicts the parameters $\mu^x$, $\Sigma_x^{-1}$ of a Gaussian mixture model p(x|y) as output. The model was and trained by maximizing the likelihood of the training data with the loss function:

$$L = \frac{1}{2} \cdot \left(x\mu_x^\top \cdot \Sigma_x^{-1} \cdot x\mu_x\right) - \log|\Sigma_x^{-1}|^{1/2}$$

**MatDesINNe framework.** The MatDesINNe framework consists of (1) training, (2) inference, (3) down-selection, and (4) localization steps. In step 1, for the INN, weight coefficients for each loss are initialized and the total loss is minimized via bi-directional training with stochastic gradient descent. The forward model with minimal l2 loss is saved as a forward surrogate. For the cINN, the loss is minimized via maximum likelihood training with stochastic gradient descent. In step 2, random samples are generated from the latent space p(z), and the corresponding posterior samples conditioned on the prior sample z and a given observation y through an invertible transformation are computed. In step 3, down-selection is performed by removing far outlier samples based on the surrogate predictions, as well as samples with parameters outside the training range. In the case of cINN, the majority of samples still remain after down-selection, while the ones which are removed are unlikely to localize to good solutions and. In step 4, localization is performed on the remaining samples by computing the gradient at the current x using the saved forward surrogate with automatic differentiation and optimized via gradient decent.

**Acknowledgements**
This work was supported by the Center for Understanding and Control of Acid Gas-Induced Evolution of Materials for Energy (UNCAGE-ME), an Energy Frontier Research Center funded by U.S. Department of Energy, Office of Science, Basic Energy Sciences. Work was performed at the Center for Nanophase Materials Sciences, which is a US Department of Energy Office of Science User Facility. VF was also supported by a Eugene P. Wigner Fellowship at Oak Ridge National Laboratory. JZ was was supported by the U.S. Department of Energy, Office of Science, Office of Advanced Scientific Computing Research, Applied Mathematics program; and by the Artificial Intelligence Initiative at the Oak Ridge National Laboratory (ORNL). ORNL is operated by UT-Battelle, LLC., for the U.S. Department of Energy under Contract DEAC05-00OR22725. This research used resources of the National Energy Research Scientific Computing Center, supported by the Office of Science of the U.S. Department of Energy under Contract No. DE-AC02-05CH11231.

**Competing Interests**
The authors declare no competing interests.
**Additional Information**
Supplementary information is available.
**Data Availability**
The DFT datasets used for training are provided at https://github.com/XXX.
**Code Availability**
The machine learning code used in this work is available at https://github.com/XXX.